\documentclass[prd,aps,preprint,nofootinbib,amssymb,eqsecnum,showkeys]{revtex4}
\usepackage{verbatim,graphics,graphicx,color,slashed}
\usepackage{ulem} 
\usepackage{hyperref}
\usepackage{changepage}

\def\lsim{\mathrel {\vcenter {\baselineskip 0pt \kern 0pt
    \hbox{$<$} \kern 0pt \hbox{$\sim$} }}}
\def\gsim{\mathrel {\vcenter {\baselineskip 0pt \kern 0pt
    \hbox{$>$} \kern 0pt \hbox{$\sim$} }}}
\def\slashchar#1{\setbox0=\hbox{$#1$}           
 \dimen0=\wd0                                 
  \setbox1=\hbox{/} \dimen1=\wd1               
\ifdim\dimen0>\dimen1                        
  \rlap{\hbox to \dimen0{\hfil/\hfil}}      
  #1                                        
  \else                                        
 \rlap{\hbox to \dimen1{\hfil$#1$\hfil}}   
   /                                         
  \fi}                                         %
\def\cpto{\mathrel {\vcenter {\baselineskip 0pt \kern 0pt
    \hbox{$CP$} \kern 0pt \hbox{$\longrightarrow$} }}}
\def\cptof{\mathrel {\vcenter {\baselineskip 0pt \kern 0pt
    \hbox{$~CP$} \kern 0pt \hbox{$\longleftrightarrow$} }}}

\begin{document}

\baselineskip=15pt

\preprint{}

\title{CP Violation in $B^0_s \to K^-\pi^+$, $B^0 \to K^+\pi^-$ Decays and \\Tests for $SU(3)$ Flavor Symmetry Predictions}

\author{Xiao-Gang He${}^{1,2,3}$\footnote{hexg@phys.ntu.edu.tw}}
\author{Siao-Fong Li$^{3}$
\footnote{r01222004@ntu.edu.tw}}
\author{Hsiu-Hsien Lin$^{3}$\footnote{r00222080@ntu.edu.tw}}
\affiliation{${}^{1}$INPAC, Department of Physics, Shanghai Jiao Tong University, Shanghai, China}
\affiliation{${}^{2}$Physics Division, National Center for Theoretical Sciences,\\
Department of Physics, National Tsing Hua University, Hsinchu, Taiwan}
\affiliation{${}^{3}$CTS, CASTS and Department of Physics, National Taiwan University, Taipei, Taiwan}

\date{\today $\vphantom{\bigg|_{\bigg|}^|}$}

\date{\today}

\vskip 1cm
\begin{abstract}
The LHCb collaboration has recently measured the first direct CP violation in $B^0_s$ decays with a rate asymmetry $A_{CP}(B^0_s\to K^- \pi^+)$ given by
$0.27 \pm 0.04(stat)\pm 0.01(syst)$. At the same time they also made the most precise measurement for $A_{CP}(B^0 \to K^+ \pi^-) = -0.080\pm 0.007(stat)\pm 0.003(syst)$.
These data confirm the predicted relation, $A_{CP}(B^0 \to K^+ \pi^- )/A_{CP}(B^0_s \to K^- \pi^+) = -  Br(B^0_s\to K^- \pi^+)\tau_{B^0}/Br(B^0 \to K^+ \pi^-)\tau_{B^0_s}$,
in the standard model with $SU(3)$ flavor symmetry. We discuss possible modifications due to $SU(3)$ breaking effects to this relation.
There are several other similar relations in B decays. Using current available data we study  whether relevant relations hold in $B^0$ and $B^0_s$ to $PP$ and $PV$ decays. Here $P$ and $V$ indicate pesudoscalar and vector mesons in the flavor octet representations.
\end{abstract}

\pacs{PACS numbers: }

\maketitle

\section{Introduction}

Decays involving a heavy b-quark have been a subject of very active research in the past decade and continue to be so at present. Data on $B$ decays from SLAC and KEK B-factory experiments Babar and Belle  provided much information  about standard model (SM), in particular in confirming SM predictions for CP violation based on Kobayashi-Maskawa (KM) mechanism~\cite{km}. Data from Tevatron confirmed many of the B-factory measurements and extended to the study of $B_s$ decays. The LHCb experiment also started to provide interesting data about $B$ and $B_s$ decays after the successful running of LHC.
Current available data for $B$ decays have been compiled by the Particle Data Group (PDG)~\cite{PDG} and also by the Heavy Flavor Average Group (HFAG)~\cite{HFAG}.
The study of $B$ and $B_s$ decays will continue to provide important information about the SM. To test the SM using B decay data sometimes have uncertainties due to our poor understanding of strong interaction at low energies. Although theoretical calculations for branching ratios have sizable uncertainties due to strong interactions, some of the SM predictions on CP violation are free from such uncertainties to high precision, such as
the mixing induced CP asymmetry in $\bar{B}^0 \to \psi K_s$
versus $B^0 \to \psi K_s$ which measures the quantity $\sin 2\beta$ in the KM unitarity triangle.  Barbar and Belle experiments have
accurately measured $\sin 2 \beta$ which provided an important information in establishing the KM mechanism for CP violation~\cite{PDG, HFAG}.  Branching ratios and CP violating rate asymmetries
for some $B$ and $B_s$ rare decays into two light mesons have also been measured~\cite{PDG,HFAG}.
These data have been used to further test the SM or to extract fundamental SM parameters.
However,  there are uncertainties due to our poor understanding of hadronic interactions~\cite{fact,qcdfact,pQCD,cdl,ylw},
one needs to find ways  to minimize these uncertainties to extract useful information.

In the lack of reliable calculations, attempts have
been made to extract useful information for the SM from symmetry considerations.
$SU(3)$ flavor symmetry is one of the symmetries which has attracted a lot of attentions. 
Here the $SU(3)$ flavor symmetry is a symmetry in QCD for strong interaction different than those additional horizontal flavor symmetries imposed 
on the SM Lagrangian for flavor physics. 
In QCD if one neglects quark masses, there is a global $SU(N)$ symmetry for $N$ quarks in which the quarks
form the fundamental representation. In practice all the existing six quarks have masses and the global $SU(N)$ is broken. In particular, the charm, beauty and top quarks have different masses larger than the QCD scale which break the symmetry badly. However,
the light quarks, $u$, $d$ and $s$ have masses smaller that the QCD scale, therefore the theory has a good approximate $SU(3)$ flavor symmetry in which the $u$, $d$ and $s$ quarks form the fundamental representation. This symmetry works well in describing low energy strong interaction phenomena. $SU(3)$ flavor symmetry has been widely used to study $B$ decays also~\cite{su3sym,su3sym1,su3sym2}.
It has been shown that using $SU(3)$ flavor symmetry for $B$ decays, it is possible to
predict many interesting relations between
CP violating observables in the SM, such as rate differences  between $\Delta S = 0$ and
$\Delta S = -1$ of $B$ or $B_s$ decay into two light mesons. These relations can provide important test for the SM.
An interesting prediction is~\cite{dh1} $\Delta(B^0\to \pi^+\pi^-) \approx - \Delta(B^0\to K^+ \pi^-)$ when small annihilation contributions are neglected. Here $\Delta(B \to PP)$ is defined as $\Delta(B\to PP) = \Gamma(\bar B\to \bar P \bar P) - \Gamma(B \to  P P)$. Experimental data are consistent with this prediction.  $SU(3)$ flavor symmetry also predict several other such relations even when annihilation contributions are kept~\cite{he1,fleischer,rosner,lipkin}.
One of them is $\Delta(B^0\to K^+\pi^-) = - \Delta(B^0_s\to K^- \pi^+)$. This relation was first studied in 1998 in Ref.\cite{he1},  later in 2000 in Ref.\cite{rosner} and in 2005 in Ref.\cite{lipkin}.

The LHCb collaboration has recently measured the first direct CP violation in $B^0_s$ decays to a better than $5\sigma$ precision with a rate asymmetry $A_{CP}(B^0_s\to K^- \pi^+)$ given by~\cite{LHCb}
$0.27 \pm 0.04(stat)\pm 0.01(syst)$. At the same time they also made the most precise measurement for $A_{CP}(B^0 \to K^+ \pi^-)$ with a value given by $ -0.080\pm 0.007(stat)\pm 0.003(syst)$.
These data confirm the predicted relation, $A_{CP}(B^0 \to K^+ \pi^- )/A_{CP}(B^0_s \to K^- \pi^+) = -  Br(B^0_s\to K^- \pi^+)\tau_{B^0}/Br(B^0 \to K^+ \pi^-)\tau_{B^0_s}$ equivalent to  $\Delta(B^0\to K^+\pi^-) = - \Delta(B^0_s\to K^- \pi^+)$, to good precision. This motivated us to revisit the $SU(3)$ prediction for this relation and to check further if there are significant  $SU(3)$ breaking effects on $\Delta(B^0\to K^+\pi^-) = - \Delta(B^0_s\to K^- \pi^+)$. We also revisit several other similar relations in $B$ and $B_s$ decays and compare them with available data.

\section{$SU(3)$ prediction for  CP asymmetry in $B^0\to K^+ \pi^-$ and $B^0_s \to K^- \pi^+$}

We start with a brief review on the derivation of $\Delta(B^0\to K^+\pi^-) = - \Delta(B^0_s\to K^- \pi^+)$ based on $SU(3)$ flavor symmetry.
The leading quark level effective Hamiltonian up to one loop level in
electroweak interaction
for hadronic charmless $B$ decays in the SM can be written as
\begin{eqnarray}
 H_{eff}^q = {4 G_{F} \over \sqrt{2}} [V_{ub}V^{*}_{uq} (c_1 O_1 +c_2 O_2)
   - \sum_{i=3}^{12}(V_{ub}V^{*}_{uq}c_{i}^{uc} +V_{tb}V_{tq}^*
   c_i^{tc})O_{i}],
\end{eqnarray}
where the coefficients
$c_{1,2}$ and $c_i^{jk}=c_i^j-c_i^k$, with $j$ indicates the internal quark,
are the Wilson Coefficients (WC). These
WC's have been evaluated by several groups~\cite{Buchalla:1995vs}. $V_{ij}$ are the KM matrix elements.
In the above the factor $V_{cb}V_{cq}^*$ has
been eliminated using the unitarity property of the KM matrix.
The operators $O_i$ are given by
\begin{eqnarray}
\begin{array}{ll}
O_1=(\bar q_i u_j)_{V-A}(\bar u_i b_j)_{V-A}\;, &
O_2=(\bar q u)_{V-A}(\bar u b)_{V-A}\;,\\
O_{3,5}=(\bar q b)_{V-A} \sum _{q'} (\bar q' q')_{V \mp A}\;,&
O_{4,6}=(\bar q_i b_j)_{V-A} \sum _{q'} (\bar q'_j q'_i)_{V \mp A}\;,\\
O_{7,9}={ 3 \over 2} (\bar q b)_{V-A} \sum _{q'} e_{q'} (\bar q' q')_{V \pm A}\;,\hspace{0.3in} &
O_{8,10}={ 3 \over 2} (\bar q_i b_j)_{V-A} \sum _{q'} e_{q'} (\bar q'_j q'_i)_{V \pm A}\;,\\
O_{11}={g_s\over 16\pi^2}\bar q \sigma_{\mu\nu} G^{\mu\nu} (1+\gamma_5)b\;,&
O_{12}={Q_b e\over 16\pi^2}\bar q \sigma_{\mu\nu} F^{\mu\nu} (1+\gamma_5)b.
\end{array}
\end{eqnarray}
where $(\bar a b)_{V-A} = \bar a \gamma_\mu (1-\gamma_5) b$, $G^{\mu\nu}$ and
$F^{\mu\nu}$ are the field strengths of the gluon and photon, respectively.

At the hadronic level, the decay amplitude can be generically written as
\begin{eqnarray}
A = <final\;state|H_{eff}^q|\bar {B}> = V_{ub}V^*_{uq} T(q) + V_{tb}V^*_{tq}P(q)\;,
\end{eqnarray}
where $T(q)$ contains contributions from
the $tree$ as well as $penguin$ due to charm and up
quark loop corrections to the matrix elements,
while $P(q)$ contains contributions purely from
one loop $penguin$ contribution. $\bar {B}$ indicates one of the $B^-$, $\bar {B}^0_d$ and $\bar {B}^0_s$ which form an 
$S(3)$ triplet.

The $SU(3)$ flavor symmetry transformation properties for operators $O_{1,2}$, $O_{3-6, 11,12}$, and $O_{7-10}$ are: $\bar 3_a + \bar 3_b +6 + \overline {15}$,
$\bar 3$, and $\bar 3_a + \bar 3_b +6 + \overline {15}$, respectively.
We indicate these representations by matrices in $SU(3)$ flavor space by $H(\bar 3)$, $H(6)$ and $H(\overline{15})$.
For $q=d$, the non-zero entries of the matrices $H(i)$ are given by~\cite{dh1,he1}
\begin{eqnarray}
H(\bar 3)^2 &=& 1\;,\;\;
H(6)^{12}_1 = H(6)^{23}_3 = 1\;,\;\;H(6)^{21}_1 = H(6)^{32}_3 =
-1\;,\nonumber\\
H(\overline {15} )^{12}_1 &=& H(\overline {15} )^{21}_1 = 3\;,\; H(\overline
{15} )^{22}_2 =
-2\;,\;
H(\overline {15} )^{32}_3 = H(\overline {15} )^{23}_3 = -1\;.
\end{eqnarray}
And for $q = s$, the non-zero entries are
\begin{eqnarray}
H(\bar 3)^3 &=& 1\;,\;\;
H(6)^{13}_1 = H(6)^{32}_2 = 1\;,\;\;H(6)^{31}_1 = H(6)^{23}_2 =
-1\;,\nonumber\\
H(\overline {15} )^{13}_1 &=& H(\overline {15} ) ^{31}_1 = 3\;,\; H(\overline
{15} )^{33}_3 =
-2\;,\;
H(\overline {15} )^{32}_2 = H(\overline {15} )^{23}_2 = -1\;.
\end{eqnarray}

These properties enable us to
write the decay amplitudes for $B$ decays into a pair of pseudoscalars $PP$ in the octet  $M=(M_{ij})$ in only a few $SU(3)$ invariant
amplitudes.

For the $T(q)$ amplitude, for example, we have
\begin{eqnarray}
T(q)&=& A_{\bar 3}^T  {B}_i H(\bar 3)^i (M_l^k M_k^l) + C^T_{\bar 3}
{B}_i M^i_kM^k_jH(\bar 3)^j \nonumber\\
&+& A^T_{6} {B}_i H(6)^{ij}_k M^l_jM^k_l + C^T_{6} {B}_iM^i_jH(6
)^{jk}_lM^l_k\nonumber\\
&+&A^T_{\overline {15}} {B}_i H(\overline {15})^{ij}_k M^l_jM^k_l +
C^T_{\overline
{15}}  {B}_iM^i_j
H(\overline {15} )^{jk}_lM^l_k\;, \label{su3}
\end{eqnarray}
where $B_i = (B^+,  B^0,  B^0_s)$
is an $SU(3)$ triplet, and $M_{i}^j$ is the $SU(3)$ pseudoscalar
octet,
\begin{eqnarray}
M= \left ( \begin{array}{ccc}
{\pi^0\over \sqrt{2}} + {\eta_8\over \sqrt{6}}&\pi^+&K^+\\
\pi^-&-{\pi^0\over \sqrt{2}} + {\eta_8\over \sqrt{6}}& K^0\\
K^-&\bar K^0&-{2\eta_8\over \sqrt{6}}
\end{array} \right ).
\end{eqnarray}

The coefficients $A_i$ and $C_i$ are constants which contain the Wilson coefficients and information about QCD dynamics. Due to the anti-symmetric property of $H(6)$ in exchanging the upper two indices,
$A_6$ and $C_6$ are not independent. For individual decay
amplitude, $A_6$ and $C_6$  always appear together in the form $C_6-A_6$.
We will absorb $A_6$ in the definition of $C_6$.
The amplitudes for $P(q)$ in terms of $SU(3)$
invariant amplitudes can be obtained in a similar way. We will indicate
the corresponding amplitudes by $A^P_i$ and $C^P_i$.

In terms of the $SU(3)$ invariant amplitudes, the decay amplitudes  for
$\bar {B}^0 \to K^- \pi^+$ and $\bar {B}^0_s \to K^+ \pi^-$ and their corresponding $B^0$ and $B^0_s$ can be written as
\begin{eqnarray}
A(\bar {B}^0 \to K^- \pi^+) &=&  V_{ub}V^*_{us} T + V_{tb}V^*_{ts} P\;,\;\;A(B^0 \to K^+ \pi^-) = V^*_{ub}V_{us} T + V^*_{tb}V_{ts} P\;,\nonumber\\
 A(\bar {B}^0_s \to K^+ \pi^-) &=&  V_{ub}V^*_{ud} T + V_{tb}V^*_{td}P\;,\;\;A(B^0_s \to K^- \pi^+) =  V^*_{ub}V_{ud} T + V^*_{tb}V_{td}P\;.
\end{eqnarray}
with
\begin{eqnarray}
T &=&C^T_{\bar 3}
 + C^T_{6} -  A^T_{\overline {15}} +  3C^T_{\overline {15} }\;,\;\;P = C^P_{\bar 3}
 + C^P_{6} - A^P_{\overline {15}} +  3C^P_{\overline {15}}.
 \end{eqnarray}

Because the KM matrix elements involved are different, the resulting decay widths for the above modes are different. However there is a relation for rate differences defined by
\begin{eqnarray}
\Delta(B\to PP) &=& \Gamma(\bar {B}\to \bar P\;\bar P) - \Gamma(B \to P \; P) \nonumber\\
&=& {\lambda_{ab}\over 8 \pi m_B} (|A(\bar B \to \bar P\;\bar P)|^2-| {A}(B \to P\; P)|^2)\;,
\end{eqnarray}
where $\lambda^B_{ab} = \sqrt{1-2(m_a^2+m_b^2)/m_B^2 + (m_a^2-m_b^2)^2/m_B^4}$ with
$m_{a,b}$ being the masses of the two particles $PP$ in the final state.

For $B^0\to K^+\pi^-$ and $B^0_s \to K^- \pi^+$ decays,  because a simple property
of the KM matrix element~\cite{Jarlskog:1985ht}, $Im (V_{ub}V_{ud}^*V_{tb}^*V_{td})
=-Im(V_{ub}V_{us}^*V_{tb}^*V_{ts})$, in the $SU(3)$ limit   we have,
\begin{eqnarray}
\Delta(B^0 \to K^+\pi^-) =- \Delta(B^0_s\to K^-\pi^+)\;.
\end{eqnarray}

Using the definition of CP asymmetry $A_{CP}(B\to PP)$,
\begin{eqnarray}
A_{CP}(B\to PP) = {\Gamma(\bar {B}\to \bar P\; \bar P) - \Gamma(B \to P \; P)\over \Gamma(\bar {B}\to \bar P\; \bar P) + \Gamma(B \to P \; P)}\;,
\end{eqnarray}
we obtain
\begin{eqnarray}
{A_{CP} (B^0\to K^+\pi^-)\over A_{CP}(B^0_s \to K^- \pi^+)} +{ Br(B^0_s\to K^- \pi^+)\tau_{B^0}\over Br(B^0\to K^+ \pi^-)\tau_{B^0_s}} =0\;. \label{relation}
\end{eqnarray}

Using experimental data the LHCb collaboration obtained numerically~\cite{LHCb} $-0.02\pm0.05\pm0.04$ for left hand side of the above equality. The relation in eq.(\ref{relation}) is consistent with data. This can be taken as a support for the $SU(3)$  prediction. Since the above relation cruicially depend on the SM relation $Im (V_{ub}V_{ud}^*V_{tb}^*V_{td})=-Im(V_{ub}V_{us}^*V_{tb}^*V_{ts})$ which holds for three generation. This can also be taken as a support for SM with three generations.

To quantify the level of whether the relation hold, we introduce a correction factor $r_{c}$ as the following
\begin{eqnarray}
{A_{CP} (B^0\to K^+\pi^-)\over A_{CP}(B^0_s \to K^- \pi^+)} +r_{c} { Br(B^0_s\to K^- \pi^+)\tau_{B^0}\over Br(B^0\to K^+ \pi^-)\tau_{B^0_s}} =0\;.  \label{correct}
\end{eqnarray}

Using LHCb data, we obtain
\begin{eqnarray}
r_{c} = 1.06\pm0.24
\end{eqnarray}
The center value is very close to the $SU(3)$ prediction, but allow breaking effects to modify the relation.

When combined available data compiled by PDG~\cite{PDG}, data from CDF~\cite{CDF}, and the recent data from LHCb~\cite{LHCb} on CP asymmetries for $B_0 \to K^+ \pi^-$ and $B_s^0\to K^- \pi^-$, we have
\begin{eqnarray}
&&A_{CP}(B^{0}\rightarrow K^{+}\pi^{-})=-0.085\pm0.006\;,\nonumber\\
&&A_{CP}(B^{0}_{s}\rightarrow K^{-}\pi^{+})=0.26\pm0.04\;.
\end{eqnarray}
The above leads to
\begin{eqnarray}
r_{c} = 1.15\pm0.22\;.
\end{eqnarray}

The central value of $r_c$ above deviated from the $SU(3)$ prediction. It is larger than that from the LHCb data alone with slightly smaller error bar. More accurate data are needed to access whether $r_c$ is really significantly away from 1 and $SU(3)$ prediction is violated.

If $r_c$ deviates from 1 will be confirmed by future experimental data to high precision, one may attribute the deviation to be due to $SU(3)$ flavor symmetry breaking. One then needs to understand how $SU(3)$ breaking effects come in to modify the relation. Unfortunately, current available methods to calculate the $B$ decay branching ratios all have large uncertainties. Still, they may provide some understandings how $SU(3)$ prediction is broken due to various effects. We briefly discuss the main features of $SU(3)$ breaking effects for the above decay modes in naive factorization, QCD factorization and pQCD calculations.

In the naive factorization calculations for the above two decays, the $SU(3)$ breaking effects come from the various meson mass differences, the $\pi$, $K$ and $B$ decay constants, and also $B \to \pi$  and $B\to K$ transition form factors. To the leading order, we have~\cite{dh1,he1}
\begin{eqnarray}
&&A(B^0 \to K^+\pi^-) \sim (m^2_{B} - m^2_\pi)f_K F_0^{B \to \pi}(m^2_K)\;,\nonumber\\
&&A(B^0_s \to K^-\pi^+) \sim (m^2_{B_s} - m^2_K)f_\pi F_0^{B \to K}(m^2_\pi)\;. \label{bbr}
\end{eqnarray}

We obtain the naive factorization approximation prediction for $r_{c}$
\begin{eqnarray}
r_{c}
&\approx & {\lambda^{B}_{K\pi}/m_{B} \over \lambda^{B_s}_{K\pi} /m_{B_s}}\left ({ (m^2_{B} - m^2_\pi)f_K F_0^{B \to \pi}(m^2_K)\over (m^2_{B_s} - m^2_K) f_\pi F_0^{B_s\to K}(m^2_\pi)} \right )^2\;.
\end{eqnarray}

In QCD factorization, the $\pi$ and $K$ meson distribution amplitude functions $\Phi_M(x)$  bring in additional $SU(3)$ breaking effects and also come from the hard scattering contributions $H^B_{M_1M_2}$~\cite{qcdfact}. The distribution amplitude $\Phi_M(x)$ is usually expanded in terms of the Gegenbauer polynomials and has
the following leading order expansion~\cite{qcdfact}
\begin{eqnarray}
\Phi_M(x) = 6x(1-x)[ 1+ \alpha_1 C^{(3/2)}_1(2x-1) +
\alpha_2C^{3/2}_2(2x-1) + ...],
\label{ge}
\end{eqnarray}
with $C^{3/2}_1(u) = 3 u$ and
$C^{3/2}_2(u) = (3/2)(5u^2-1)$,
and the coefficients $\alpha_i$ are different for $\pi$ and $K$.

Following discussions in Ref.\cite{dhv,bn1,wz} and neglecting small annihilation contributions, we obtain the leading  QCD factorization approximation for $r_{c}$,
\begin{eqnarray}
r_{c}
&\approx & {\lambda^{B}_{K\pi}/m_{B} \over \lambda^{B_s}_{K\pi} /m_{B_s}}\left ({ (m^2_{B} - m^2_\pi)f_K F_0^{B \to \pi}(m^2_K)\over (m^2_{B_s} - m^2_K) f_\pi F_0^{B_s\to K}(m^2_\pi)} \right )^2\nonumber\\
&\times&
\left[ \frac{1- 0.748 \alpha^K_1 - 0.109
\alpha^K_2 - 0.017 H^{B}_{K \pi}  }{
1- 0.748 \alpha^\pi_1  - 0.109
\alpha^\pi_2 - 0.017 H^{B_s}_{\pi K} } \right].
\label{relgmod}
\end{eqnarray}

$H^{B_s}_{K\pi}$ and $H^B_{\pi K}$ are related approximately by,
$H^{B_s}_{K\pi} \approx (f_{B_s} f_K / m_{B_s}\lambda_{B_s} F_0^{B_s\to K}(m^2_\pi))(m_{B} \lambda_{B} F^{B^0\to \pi}_0(m^2_K)/ f_{B}f_\pi)H^{B}_{\pi K}$.
$H^{B_s}_{\pi K}$ and $H^B_{K\pi}$ are in the range between 0.8 to 1~\cite{qcdfact}.

For numerical evaluation, we use  $f_{\pi}=(130.41\pm0.03\pm0.20)$ Mev, $f_{K}=(156.1\pm0.2\pm0.8\pm0.2)$ Mev, $f_{B}=186$ MeV, $f_{B_s}=224$ MeV, $F^{B\to {\pi}}=0.28^{+0.05}_{-0.02}$, $F^{B_{s}\to\pi}=0.30^{+0.04}_{-0.03}$, $\alpha_1^\pi = 0$, $\alpha^\pi_2 =0.3$, $\alpha^K_1=0.10 \pm 0.04$, $\alpha^K_2 = 0.1 \pm 0.3$, $\lambda_B=\lambda_{B_{s}}=0.350$.~\cite{PDG,values}. The range of $r_{c}$ is estimated to be in the range of 0.86 to 1.67 with a central value 1.15. The theoretical allowed range is consistent with current data.

In pQCD method, there is no need to introduce transition form factors which are parts evaluated within the method.  The $SU(3)$ breaking effects for the relation in eq.(\ref{relation}) has been studied. The resulting $r_c$ is estimated in the range of~\cite{cdl} $0.90  - 1.10$ with a central value 1.00. This is also consistent with the data allowed range.

With more precise measurements on CP violation for $B^0\to K^+\pi^-$ and $B^0_s \to K^- \pi^+$ decays, we may still have difficulties to distinguish various ways of theoretical calculations for these decays.
Since the $SU(3)$ relation of eq.\ref{relation} holds well, one wonders if similar relations predicted with $SU(3)$ symmetry also hold. In the following section we compare other similar $SU(3)$  predictions  for $B\to PP$ and $B\to VP$ decay modes with available data.

\section{CP asymmetry relations in other $B\to PP$ and $B\to VP$ decays}

Using the amplitudes obtained in eq. \ref{su3} for $T(q)$ and similar ones for $P(q)$,
one can find several other CP asymmetry relations for $B\to PP$ decays.
These relations can be used to test $SU(3)$ prediction further.  Several of the relations contains $\eta_8$ in the final states. It is well known that there is mixing between $\eta_8$ and $\eta_1$. Therefore decay modes involving $\eta$ are more complicated to analyse. We will only list and study those decays containing pions and kaons in the final states. They are~\cite{he1},
\begin{eqnarray}
P1)&\;\;&
\Delta(B^+ \to K^+ \bar K^0) = - \Delta (B^+ \to K^0 \pi^+ )\;,\nonumber\\
P2)&& \Delta(B^0 \to \pi^+ \pi^-) = - \Delta (B^0_s \to K^- K^+)\;,\nonumber\\
P3)&& \Delta(B^0 \to K^+ K^-) = - \Delta (B^0_s \to \pi^+ \pi^-) =-2\Delta(B^0_s\to \pi^0\pi^0)\;,\nonumber\\
P4)&& \Delta(B^0 \to \bar K^0 K^0) = - \Delta (B^0_s \to K^0 \bar K^0)\;,\nonumber\\
P5)&&  \Delta (B^0_s\to K^-\pi^+)=- \Delta(B^0\to K^+ \pi^-),\nonumber\\
P6)&& \Delta ( B^0_s \to \bar K^0 \pi^0)=- \Delta(B^0 \to K^0 \pi^0). \label{rpp}
\end{eqnarray}

The amplitudes $A_{\bar 3, \overline  {15}}$  correspond to annihilation
contributions.  It has been argued that these contributions are small
based on model calculations and also supported by B decay data. Processes having
$C_i$ contributions typically have branching ratio of order a few times $10^{-6}$.
The annihilation contribution induced  $B^0 \to K^+K^-$ has a branching ratio less
than~\cite{PDG,HFAG}  $4.1\times 10^{-7}$ at
90\% confidence level. This is sometimes taken as argument for the smallness of annihilation contribution.
The annihilation induced $B_s^0 \to \pi^+\pi^-$ decay has been measured to have a branching ratio~\cite{HFAG}
$(0.73\pm 0.14)\times 10^{-6}$. Although it is still small, it may lead some modifications and one should be more careful in neglecting annihilation contributions.
In the limit that annihilation contributions are small, it is difficult to
perform test for P3)
because they are all annihilation contribution induced decays.
Neglecting annihilation contributions, one has additional relations~\cite{dh1,he1}
\begin{eqnarray}
&&P1) \approx P4)\;,\;\;\;\;P2) \approx  P5)\;, \;\;\;\;P6) \approx  \Delta( B^0 \to \pi^0\pi^0)\;.\label{rpp1}
\end{eqnarray}

In a similar fashion, one can study CP violating relations for $B \to V P$ in the $SU(3)$ limit~\cite{he2}. Here $V$ is one of the
vector mesons $(\rho^\pm,\; \rho^0\;, K^{*0},\; \bar {K}^{*0},\; K^{*\pm},\; \omega,\; \phi)$. We find the following relations
exist,
\begin{eqnarray}
V1)&\;\;& \Delta(B^+ \to K^+ \bar K^{*0}) = - \Delta (B^+ \to K^{*0} \pi^+ )\;,\nonumber\\
V2)&& \Delta(B^0 \to  K^0 \bar K^{*0}) = - \Delta (B^0_s \to \bar K^0  K^{*0})\;,\nonumber\\
V3)&& \Delta(B^+ \to  \bar K^0 K^{*+}) = - \Delta (B^+ \to  K^0  \rho^+)\;,\nonumber\\
V4)&& \Delta(B^0\to \bar K^0  K^{*0}) = - \Delta (B^0_s \to  K^0 \bar K^{*0})\;,\nonumber\\
V5)&& \Delta(B^0 \to \pi^+ \rho^-) = - \Delta (B^0_s \to K^+ K^{*-})\;,\nonumber\\
V6)&& \Delta (B^0_s\to\pi^+K^{*-}) = -\Delta(B^0 \to  K^+\rho^-) \;,\nonumber\\
V7)&& \Delta(B^0 \to \pi^- \rho^+) = - \Delta (B^0_s \to K^- K^{*+})\;,\nonumber\\
V8)&& \Delta (B^0_s\to K^-\rho^+) = -\Delta(B^0\to \pi^- K^{*+})\;,\nonumber\\
V9)&& \Delta(B^0 \to K^+ K^{*-}) = - \Delta (B^0_s \to \pi^+ \rho^-)\;,\nonumber\\
V10)&& \Delta(B^0 \to K^- K^{*+}) = - \Delta (B^0_s \to \pi^- \rho^+). \label{rpv}
\end{eqnarray}

The decay modes in relations V9) and V10) are annihilation contribution induced decays and have small branching ratios.
Neglecting annihilation contribution, there are additional relations for rate differences.
We find the following approximate equalities,

\begin{eqnarray}
\begin{array}{ll}
V1)\approx V2)\;,\;\;\;\;\;V3) \approx V4)\;,\;\;\;\;V5)\approx V6)\;,\;\;\;\;V7)\approx V8).
\end{array}\label{rpv1}
\end{eqnarray}

We  collect current experimental data on related decays and some $SU(3)$ predictions in Tables \ref{t2} to \ref{t8}.
The Heavy Flavor Averaging Group (HFAG) has collected relevant data in Ref.\cite{HFAG}. Averaged data for some relevant
$B$ decays have also been collected in Ref.\cite{PDG} by the Particle Data Group (PDG). Since the HFAG data were compiled later and have included some new information which has not been taken into account in the PDG compilation, we will use the HFAG averaged data in Ref.\cite{HFAG} for discussions.
We also used the new information on CP violation in $B_0 \to K^+ \pi^-$ and $B^0_s \to K^- \pi^+$ from LHCb. So the values for these two decays are different than the HFAG values.

We have compiled the data in such a way that, in the Tables \ref{t4} and \ref{t8}, for CP asymmetry the first columns are related to $\Delta S = 0$ and the third columns are related to  $\Delta S = -1$ $B$ and $B_s$ decays which are in order according to eqs. (\ref{rpp}, \ref{rpp1}) and eqs.(\ref{rpv}, \ref{rpv1}).  Using known branching ratio and CP asymmetry for one decay mode in the relations in eqs.(\ref{rpp}$-$ \ref{rpv1}), one can predict the other's CP asymmetry when the branching ratio is also know which are shown in the second and fourth columns.

For $B\to PP$ decays, we have information to make some meaningful predictions as can be seen from Tables \ref{t2} and \ref{t4}.  We make a few comments on the results in the following.

\begin{itemize}

\item From Table \ref{t4}, we see that the signs of the central values for CP asymmetry for the two decay modes  in P1) are in agreement with $SU(3)$ prediction. But the sizes of the central values are not in agreement with predictions. However, the error bars are too large to make a conclusion.

\item For the two decay modes in P2), the CP asymmetry is $A_{CP}(B^0 \to \pi^+\pi^-)$ has been measured. The predicted CP asymmetry $A_{CP}(B^0_s\to K^-K^+)$ is different in sign and also different in size with data. However the error bar of the data is large for $A_{CP}(B^0_s\to K^+ K^-)$. One cannot conclude inconsistent here. Accurate measurement of CP asymmetry for $B^0_s\to K^- K^+$ can provide very important information about CP violation in SM and about $SU(3)$ symmetry.

\item Decay modes in P3) are annihilation contribution induced decays and all have small branching rations.  No CP asymmetry has been measured for any of them. Therefore it is not possible to check the consistence of data and theory predictions here.

\item The decay modes in relation P4) also have small branching ratios and no CP asymmetry data are available to make predictions and compare with data.

\item The CP asymmetries are best measured for the decay modes in P5) and have been discussed previously. The relation P5) agrees with data very well.

\item For P6), although there is information for the branching ratio and CP asymmetry for $B^0\to K^0 \pi^0$ decay, no information on the branching ratio or CP asymmetry for $B^0_s\to \bar K^0\pi^0$, it is not possible to make predictions for CP asymmetry at this moment.

\item When annihilation contributions are neglected, one would have: P1) $\approx$ P4), P2) $\approx$  P5), and P6) $\approx  \Delta( B^0 \to \pi^0\pi^0)$. Since no information on CP asymmetries for the modes in relation P4), we cannot check whether P1) $\approx $ P4). But we can check whether P2) $\approx$ P5). $SU(3)$ predicts $A_{CP}(B^0\to \pi^+\pi^-)$ to be approximately equal to $A_{CP}(B^0_s\to K^- \pi^+)$. Experimental data agree with this prediction very well. This can be taken as an indication that annihilation is indeed small. For the prediction P6) $\approx  \Delta( B^0 \to \pi^0\pi^0)$,  one can use the measured $A_{CP}(B^0\to \pi^0\pi^0)$ to predict $A_{CP}(B^0\to K^0 \pi^0)$. The sign of the central value for this  is in agreement with data, but the size is different. Again, the error bar for $A_{CP}(B^0\to K^0\pi^0)$ is large and cannot exclude the $SU(3)$ relation.

\end{itemize}

For $B\to PV$ decays, the available data to test the CP asymmetry relations are limited as can be seen from Tables \ref{t6} and \ref{t8}. Only approximate relations
V5)$\approx$ V6), and V7)$\approx$ V8) have data available to check. For V5)$\approx$ V6), one can use data for $B^0\to \pi^+ \rho^-$
to predict CP asymmetry for $B^0\to K^+ \rho^-$ or vice versa. The predicted central value is in agreement with data in sign, but the sizes are different.   For V7)$\approx$ V8), one can use $B^0\to \pi^-\rho^+$ data to predict CP asymmetry for $B^0\to \pi^- K^{*+}$. This time the central values not only sizes are different, but also the signs are different than predictions. There seems to be a difference at about 4$\sigma$ level. Of course this needs to be further confirmed by experimental data. However, we also note that the branching ratios and CP asymmetries are for the average of $B^0\to \pi^+ \rho^-$ and $B^0\to \pi^- \rho^+$. These two decay modes in general have different branching ratios and CP asymmetries. The averaged branching ratios and CP asymmetries
may not satisfy the $SU(3)$ relations discussed here~\cite{he2}. One needs to separately measure the branching ratios and CP asymmetries for these decay modes. To achieve this, one needs to tag $B^0$ and $\bar B^0$ for each decay. Before these have been done, one cannot truly test the relations under consideration.

From the above discussions, we see that at present available data are consistent with $SU(3)$ predictions. When more data become available, these B decay modes will provide important information about the SM and also $SU(3)$ flavor symmetry in B decays.

\section{Summary}

The LHCb collaboration has measured CP violating observables  $A_{CP}(B^0_s\to K^- \pi^+)$ and $A_{CP}(B^0 \to K^+ \pi^-)$ to high precision. These data have shown that one of the $SU(3)$ flavor symmetry predicted relation $A_{CP}(B^0 \to K^+ \pi^- )/A_{CP}(B^0_s \to K^- \pi^+) = -  Br(B^0_s\to K^- \pi^+)\tau_{B^0}/Br(B^0 \to K^+ \pi^-)\tau_{B^0_s}$ is well respected.  When the LHCb results are combined with Babar, Belle and Tevatron data, we find that the central value deviate from $SU(3)$ prediction to be larger than that from the LHCb data alone. We studied possible $SU(3)$ breaking effects in these decays using naive factorization, QCD factorization and pQCD method. There are different sources which break the $SU(3)$ flavor symmetry, such as the meson decay constants, the B to light meson transition form factors, and also in distribution functions for meson amplitudes. We however found that these different methods give ranges which cover the $SU(3)$  predictions and the current experimental data allowed ranges.

With current data, it is not possible to conclude that large $SU(3)$ breaking exists in the relation under consideration. $SU(3)$ relations for CP asymmetries may well hold. This motivated us to revisit $SU(3)$ predictions for CP asymmetry in $B$ decays. For $B\to PP$ decays, current available information allow us to make some meaningful predictions.
The CP asymmetries are best measured for the decay modes in P5). They agree with $SU(3)$ prediction very well.
The signs of the central values for CP asymmetry for the two decay modes  in P1) are in agreement with $SU(3)$ prediction. But the sizes of the central values are not in agreement with predictions.
For the two decay modes in P2), the CP asymmetry is $A_{CP}(B^0 \to \pi^+\pi^-)$ has been measured. The predicted CP asymmetry $A_{CP}(B^0_s\to K^-K^+)$ is different in sign and also different in size with data. However the error bar of the data is large for $A_{CP}(B^0_s\to K^+ K^-)$.  At present it is not possible to rule out the theoretical relations. Future improved data can provide crucial information to test the relations.

When annihilation contributions are neglected, $SU(3)$ predicts $A_{CP}(B^0\to \pi^+\pi^-)$ to be approximately equal to $A_{CP}(B^0_s\to K^- \pi^+)$. Experimental data agree with this prediction very well. This can be taken as an indication that annihilation is indeed small. For the prediction P6) $\approx  \Delta( B^0 \to \pi^0\pi^0)$,  one can use the measured $A_{CP}(B^0\to \pi^0\pi^0)$ to predict $A_{CP}(B^0\to K^0 \pi^0)$. The sign of the central value for this  is in agreement with data, but the size is different. Again, the error bar for $A_{CP}(B^0\to K^0\pi^0)$ is large and cannot exclude the $SU(3)$ relation.

For $B\to PV$ decays, the available data to test the CP asymmetry relations are limited. Only approximate relations
V5)$\approx$ V6), and V7)$\approx$ V8) have data available to test. For V5)$\approx$ V6), one can use data for $B^0\to \pi^+ \rho^-$ to predict CP asymmetry for $B^0\to K^+ \rho^-$ or vice versa. The predicted central value is in agreement with data in sign, but the sizes are different.   For V7)$\approx$ V8), one can use $B^0\to \pi^-\rho^+$ data to predict CP asymmetry for $B^0\to \pi^- K^{*+}$. This time the central values not only sizes are different, but also the signs are different than predictions. However, one should note that the branching ratios and CP asymmetries compiled by HFAG are for the average of $B^0\to \pi^+ \rho^-$ and $B^0\to \pi^- \rho^+$. These two decay modes in general have different branching ratios and CP asymmetries. The averaged branching ratios and CP asymmetries
may not satisfy the $SU(3)$ relations discussed. One needs to separately measure the branching ratios and CP asymmetries for these decay modes. To achieve this, one needs to tag $B^0$ and $\bar B^0$ for each decay. Before these have been done, one cannot truly test the relations.

We conclude that the present data have provided some information about $SU(3)$ predictions for CP asymmetries. The well measured decays modes agree with $SU(3)$ predictions. When more data become available several other CP asymmetry relations predicted in the SM with $SU(3)$ symmetry can be  tested.

\begin{acknowledgments}

The work was supported in part by MOE Academic Excellent Program (Grant No: 102R891505) and NSC of ROC, and in part by NNSF(Grant No:11175115) and Shanghai science and technology commission (Grant No: 11DZ2260700) of PRC.

\end{acknowledgments}

\begin{table}[h]\footnotesize
\caption{The experimental results for $Br(B\rightarrow PP)$ from HFAG. The sign ``$- -$'' indicates that no information is available for the relevant decays.}
\begin{tabular}{|c|c|c|c|c|c|c|}
\hline
&$\Delta S =0\ Process$  &$Br^{HFAG}(10^{-6})$& $\Delta S=-1\  Process$ &$Br^{HFAG}(10^{-6})$ \\
\hline
P1)&$B^{+}\rightarrow K^{+}\overline{K}^{0}$  &$1.19\pm 0.18$& $B^{+}\rightarrow K^{0}\pi^{+}$  &$23.80\pm 0.74$\\
\hline
P2)&$B^{0}\rightarrow \pi^{+}\pi^{-}$ &$5.10\pm 0.19$&
$B^{0}_{s}\rightarrow K^{-}K^{+}$ &$24.5\pm 1.8$ \\
\hline
P3)&$B^{0}\rightarrow K^{+}K^{-}$ & $0.12\pm0.06$ &
$B^{0}_{s}\rightarrow \pi^{+}\pi^{-}$ & $0.73\pm0.14$  \\
&&&$B^{0}_{s}\rightarrow\pi^{0}\pi^{0}$& $ --$\\
\hline
P4)&$B^{0}\rightarrow \overline{K}^{0}K^{0}$ &$1.21\pm0.16$& $B^{0}_{s}\rightarrow K^{0}\overline{K}^{0}$ &$<66$ \\
\hline
P5)&$B^{0}_{s}\rightarrow K^{-}\pi^{+}$ &$5.4\pm 0.6$&
$B^{0}\rightarrow K^{+}\pi^{-}$ &$19.55^{+0.54}_{-0.53}$ \\
\hline
P6)&$B^{0}_{s}\rightarrow \overline{K}^{0}\pi^{0}$ &$--$ &
$B^{0}\rightarrow K^{0}\pi^{0}$ &$9.92^{+0.49}_{-0.48}$ \\
&&&$B^{0}\rightarrow\pi^{0}\pi^{0}$& $1.91^{+0.22}_{-0.23}$  \\
\hline
\end{tabular}\label{t2}
\end{table}

\begin{table}[h]\footnotesize
\caption{The experimental and predicted results for $A_{CP}(B\rightarrow PP)$ from HFAG. The sign ``$- -$'' indicates that no information is available for the relevant decays.}
\begin{tabular}{|c|c|c|c|c|c|c|}
\hline
&$A_{CP}^{Exp.}(\Delta S =0)$&$A_{CP}^{Pred.}(\Delta S=-1)$&
$A_{CP}^{Exp.}(\Delta S=-1)$&$A_{CP}^{Pred.}(\Delta S=0)$ \\
\hline
P1)&$B^{+}\rightarrow K^{+}\overline{K}^{0}$ &$B^{+}\rightarrow K^{0}\pi^{+}$&
$B^{+}\rightarrow K^{0}\pi^{+}$ &$B^{+}\rightarrow K^{+}\overline{K}^{0}$\\
&$0.041\pm0.141$ & $-0.0021\pm0.0071$ &   $-0.015\pm 0.012$ & $0.300\pm0.244$ \\

\hline
P2)&$B^{0}\rightarrow \pi^{+}\pi^{-}$ &$B^{0}_{s}\rightarrow K^{-}K^{+}$&
$B^{0}_{s}\rightarrow K^{-}K^{+}$ &$B^{0}\rightarrow \pi^{+}\pi^{-}$\\
&$0.29\pm0.05$ & $-0.060\pm0.011$  & $0.02\pm0.18\pm0.04$  & $-0.097\pm0.892$ \\

\hline
P3)&$B^{0}\rightarrow K^{+}K^{-}$  &$B^{0}_{s}\rightarrow \pi^{+}\pi^{-}$&
$B^{0}_{s}\rightarrow \pi^{+}\pi^{-}$  &$B^{0}\rightarrow K^{+}K^{-}$\\
&$--$&&$--$&\\
 &&$B^{0}_{s}\rightarrow\pi^{0}\pi^{0}$ & $B^{0}_{s}\rightarrow\pi^{0}\pi^{0}$& $B^{0}\rightarrow K^{+}K^{-}$  \\
&&&$--$&\\

\hline
P4)&$B^{0}\rightarrow \overline{K}^{0}K^{0}$ &$B^{0}_{s}\rightarrow K^{0}\overline{K}^{0}$&
$B^{0}_{s}\rightarrow K^{0}\overline{K}^{0}$ &$B^{0}\rightarrow \overline{K}^{0}K^{0}$\\
&$--$&&$--$&\\

\hline
P5)&$B^{0}_{s}\rightarrow K^{-}\pi^{+}$ &$B^{0}\rightarrow K^{+}\pi^{-}$&
$B^{0}\rightarrow K^{+}\pi^{-}$ &$B^{0}_{s}\rightarrow K^{-}\pi^{+}$\\
&$0.26\pm0.04$ & $-0.073\pm0.010$ & $-0.085\pm0.006$ & $0.304\pm0.040$ \\

\hline
P6)&$B^{0}_{s}\rightarrow \overline{K}^{0}\pi^{0}$  &$B^{0}\rightarrow K^{0}\pi^{0}$& $B^{0}\rightarrow K^{0}\pi^{0}$ &$B^{0}_s\rightarrow \overline{K}^{0}\pi^{0}$\\
&$--$ &  & $-0.01\pm0.10$  &  \\
&$B^0\to \pi^0\pi^0$&$B^{0}\rightarrow K^{0}\pi^{0}$ &&$B^0\to \pi^0\pi^0$\\
&$0.43\pm0.24$  &$-0.083\pm 0.047$  & &$0.052\pm 0.519$\\
\hline
\end{tabular}\label{t4}
\end{table}

\begin{table}[h]\footnotesize
\caption{The experimental results for $Br(B\rightarrow PV)$ from HFAG. The sign ``$- -$'' indicates that no information is available for the relevant decays.}
\begin{tabular}{|c|c|c|c|c|}
\hline
&$\Delta S =0\ Process$  &$Br^{HFAG}(10^{-6})$& $\Delta S=-1\  Process$  &$Br^{HFAG}(10^{-6})$ \\
\hline
V1)&$B^+ \to K^+ \bar K^{*0}$& $--$ &$B^+ \to K^{*0} \pi^+$ &$9.9^{+0.8}_{-0.9}$  \\
\hline
V2)&$B^0 \to  K^0 \bar K^{*0}$& $--$ &$B^0_s \to \bar K^0  K^{*0}$& $--$ \\
\hline
V3)&$B^+ \to \bar K^0 K^{*+}$& $--$ & $B^+ \to K^0  \rho^+$&$8.0^{+1.5}_{-1.4}$ \\
\hline
V4)&$B^0\to \bar K^0  K^{*0}$&$<1.9$ &$B^0_s \to  K^0 \bar K^{*0}$& $--$ \\
\hline
V5)&$B^0 \to \pi^+ \rho^-$&$23\pm 2.3$&$B^0_s \to K^+ K^{*-}$& $--$ \\
\hline
V6)&$B^0_s\to\pi^+K^{*-}$& $--$ &$B^0 \to  K^+\rho^-$&$7.2\pm0.9$\\
\hline
V7)&$B^0 \to \pi^- \rho^+$&$23\pm 2.3$&$B^0_s \to K^- K^{*+}$& $--$   \\
\hline
V8)&$B^0_s\to K^-\rho^+$& $--$ &
$B^0\to \pi^- K^{*+}$&$8.5\pm0.7$\\
\hline
V9)&$B^0 \to K^+ K^{*-}$& $--$ &$B^0_s \to \pi^+ \rho^-$&  $--$  \\
\hline
V10)&$B^0 \to K^- K^{*+}$& $--$ &$B^0_s \to \pi^- \rho^+$& $--$   \\
\hline
\end{tabular}\label{t6}
\end{table}

\begin{table}[!]\footnotesize
\caption{The experimental and predicted results for $A_{CP}(B\rightarrow PV)$ from HFAG. The sign ``$- -$'' indicates that no information is available for the relevant decays.}
\begin{tabular}{|c|c|c|c|c|}
\hline
&$A_{CP}^{Exp.}(\Delta S =0)$&$A_{CP}^{Pred.}(\Delta S=-1)$&
$A_{CP}^{Exp.}(\Delta S=-1)$&$A_{CP}^{Pred.}(\Delta S=0)$ \\

\hline
V1)&$B^+ \to K^+ \bar K^{*0}$&$B^+ \to K^{*0} \pi^+$&
$B^+ \to K^{*0} \pi^+$&$B^+ \to K^+ \bar K^{*0}$\\
& $--$ &~&$-0.038\pm 0.042$&~\\

\hline
V2)&$B^0 \to  K^0 \bar K^{*0}$&$B^0_s \to \bar K^0  K^{*0}$&
$B^0_s \to \bar K^0  K^{*0}$&$B^0 \to  K^0 \bar K^{*0}$\\
& $--$ &~& $--$ &~\\

\hline
V3)&$B^+ \to \bar K^0 K^{*+}$ & $B^+ \to \bar K^0 K^{*+}$&
$B^+ \to K^0  \rho^+$&$B^+ \to \bar K^0 K^{*+}$\\
& $--$ &~&$-0.12\pm0.17$&~\\

\hline
V4)&$B^0\to \bar K^0  K^{*0}$& $B^0_s \to  K^0 \bar K^{*0}$ &
$B^0_s \to  K^0 \bar K^{*0}$&$B^0\to \bar K^0  K^{*0}$ \\
& $--$ &~& $--$ &~\\

\hline
V5)&$B^0 \to \pi^+ \rho^-$ &$B^0_s \to K^+ K^{*-}$&
$B^0_s \to K^+ K^{*-}$&$B^0 \to \pi^+ \rho^-$\\
&$-0.13\pm0.04$&~& $--$ &~\\
&~&$ B^0 \to  K^+\rho^-$&~&~\\
&~&$0.415\pm0.144$&~&~\\

\hline
V6)&$B^0_s\to\pi^+K^{*-}$&$B^0 \to  K^+\rho^-$&
$B^0 \to  K^+\rho^-$&$B^0_s\to\pi^+K^{*-}$\\
& $--$ &~&$0.20\pm0.11$&~\\
&~&~&~&$B^0 \to \pi^+ \rho^-$\\
&~&~&~&$-0.063\pm0.036$\\

\hline
V7)&$B^0 \to \pi^- \rho^+$&$B^0_s \to K^- K^{*+}$&
$B^0_s \to K^- K^{*+}$&$B^0 \to \pi^- \rho^+$\\
&$-0.13\pm0.04$&~& $--$ &~\\
&~&$B^0\to \pi^- K^{*+}$&~&~\\
&~&$0.352\pm0.117$&~&~\\

\hline
V8)&$B^0_s\to K^-\rho^+$&$B^0\to \pi^- K^{*+}$&
$B^0\to \pi^- K^{*+}$&$B^0_s\to K^-\rho^+$\\
& $--$ &~&$-0.23\pm0.06$ &~\\
&~&~&~&$B^0 \to \pi^- \rho^+$\\
&~&~&~&$0.085\pm0.025$\\

\hline
V9)&$B^0 \to K^+ K^{*-}$& $B^0_s \to \pi^+ \rho^-$ &
$B^0_s \to \pi^+ \rho^-$&$B^0 \to K^+ K^{*-}$\\
& $--$ &~& $--$ &~\\

\hline
V10)&$B^0 \to K^- K^{*+}$ & $B^0_s \to \pi^- \rho^+$ &
$B^0_s \to \pi^- \rho^+$&$B^0 \to K^- K^{*+}$\\
& $--$ &~& $--$ &~\\

\hline
\end{tabular}\label{t8}
\end{table}

\end{document}